\documentstyle[NATO,numreferences,epsfig]{CRCKAPB} 

\newcommand{\timav}{\langle \dot M \rangle}

\begin{opening}
\title{INTERNAL CONDITIONS OF ACCRETING WHITE DWARFS}

\author{LARS BILDSTEN}
\institute{Kavli Institute for Theoretical Physics and Department of Physics, University of California, Santa Barbara\\Kohn Hall, Santa Barbara, CA 93106}
\author{DEAN M. TOWNSLEY}
\institute{Department of Physics, University of California, Santa Barbara Broida Hall, Santa Barbara, CA 93106}

\end{opening}

\begin{document}
\begin{abstract}

 We explain the physics of compressional heating of the deep interior
of an accreting white dwarf (WD) at accretion rates low enough so that
the accumulated hydrogen burns unstably and initiates a classical nova
(CN).  In this limit, the WD core temperature ($T_c$) reaches an
equilibrium value ($T_{\rm c,eq}$) after accreting an amount of mass much
less than the WD's mass. Once this equilibrium is reached, the
compressional heating from within the envelope exits the surface. This
equilibrium yields useful relations between the WD surface
temperature, accretion rate and mass that can be employed to measure
accretion rates from observed WD effective temperatures, thus testing
binary evolution models for cataclysmic variables.
 
\end{abstract}

\section{Overview and Context} 

 Cataclysmic variables (CV) are formed when the WD made
during a common envelope event finally comes into contact with its
companion as a result of gravitational radiation losses
over a few Gyr (see \cite{how}). The WD will have cooled during this
time; a $0.20 M_\odot$ He WD would have $T_c=3.3\times
10^6 \ {\rm K} $ at 4 Gyr \cite{alt}, whereas a $0.6M_\odot$ C/O
WD would have $T_c=2.5\times 10^6 {\rm K}$ in 4 Gyr \cite{sal}. These
give effective temperatures $T_{\rm eff}\approx 4500-5000$ K.

A subset of the CVs, called Dwarf Novae (DN), contain a WD accreting
at low time-averaged rates $\timav<10^{-9}M_\odot \ {\rm yr}^{-1}$,
where the accretion disk is subject to a thermal instability which
causes it to rapidly transfer matter onto the WD (at $\dot M \gg
\timav$) for a week once every month to year. The $\dot M$ onto the WD
is often low enough between outbursts that the UV emission is
dominated by the internal luminosity of the WD, allowing for a
measurement of the WD luminosity, yielding $T_{\rm eff}>10,000\ {\rm
K}$ \cite{sion}.

 The WD is clearly hotter than expected for its age, providing 
evidence of the thermal impact of prolonged accretion \cite{sion}. The
theoretical illumination of this phenomena will aid our understanding
of CV evolution, ejection of material during CN events, and finally
allow for theoretical seismology of the one known non-radial pulsator,
GW Lib \cite{zyl,szkody}. We summarize here both our published work 
\cite{tow02,towbil} and 
work in progress.

\section{Compressional and Nuclear Heating in the White Dwarf}
 
At low $\timav$'s, the energy from accretion onto the WD
surface is radiated away near the photosphere and does not penetrate
the interior. We are thus concerned about energy released deep within
the WD, far beneath the photosphere.  Compressional heating is the
energy released by fluid elements as they are buried by further
accretion.  Energy is released because the time it takes a fluid
element to move inward is much longer than the time for heat
transport. Hence, the fluid element loses entropy as it moved inward,
 adjusting
its temperature to the profile needed to carry the exiting flux.

An estimate of the energy released this way is just the gravitational
energy liberated as a fluid element moves down in the WD's
gravitational field, $g=GM/R^2$.  In the outer atmosphere, a fluid
element moves a scale height, $h=kT/\mu m_p g$, in the time it takes
to replace it by accretion, giving $L\sim \timav g h\sim \timav k
T/\mu m_p$. This has the correct scaling, notably the dependence
on $\mu$ which is a contrasting parameter between the accreted H/He
envelope and the C/O core. Our previous work \cite{tow02}
showed that the dominant energy release is in the accreted
outer envelope, giving $L\approx 3kT_c \timav/\mu_e m_p$, where
$\mu_e\approx 0.6$ is the mean molecular weight of the accreted
material.

An additional energy source is slow nuclear ``simmering'' near the
base of the accreted layer.  This is significant when the layer
becomes thick, eventually becoming thermally unstable to ignite the
CN. The CN energy is assumed to be radiated away in the
explosion. However, the nuclear simmering prior to the CN releases an
amount of energy comparable to that from compression.  Because of
unstable nuclear burning and the resulting CN cycle, the H/He envelope
mass changes with time. This allows the core to cool at low
accumulated masses and be heated prior to unstable ignition
\cite{tow02}. We use CN ignition to determine the maximum mass of
the overlying freshly accreted shell, and find the steady-state (i.e.
cooling equals heating throughout the CN cycle) core temperature,
$T_{\rm c,eq}$, as a function of $\timav$ and $M$. The left panel in
Figure 1 shows the resulting $T_{\rm c,eq}$ and maximum
accumulated mass as a function of $\timav$. These masses should be
compared to measured ejection masses in order to learn if the WD is
losing or gaining mass during CN outburst. 

\begin{figure}
\epsfxsize=2.32in
\epsfbox{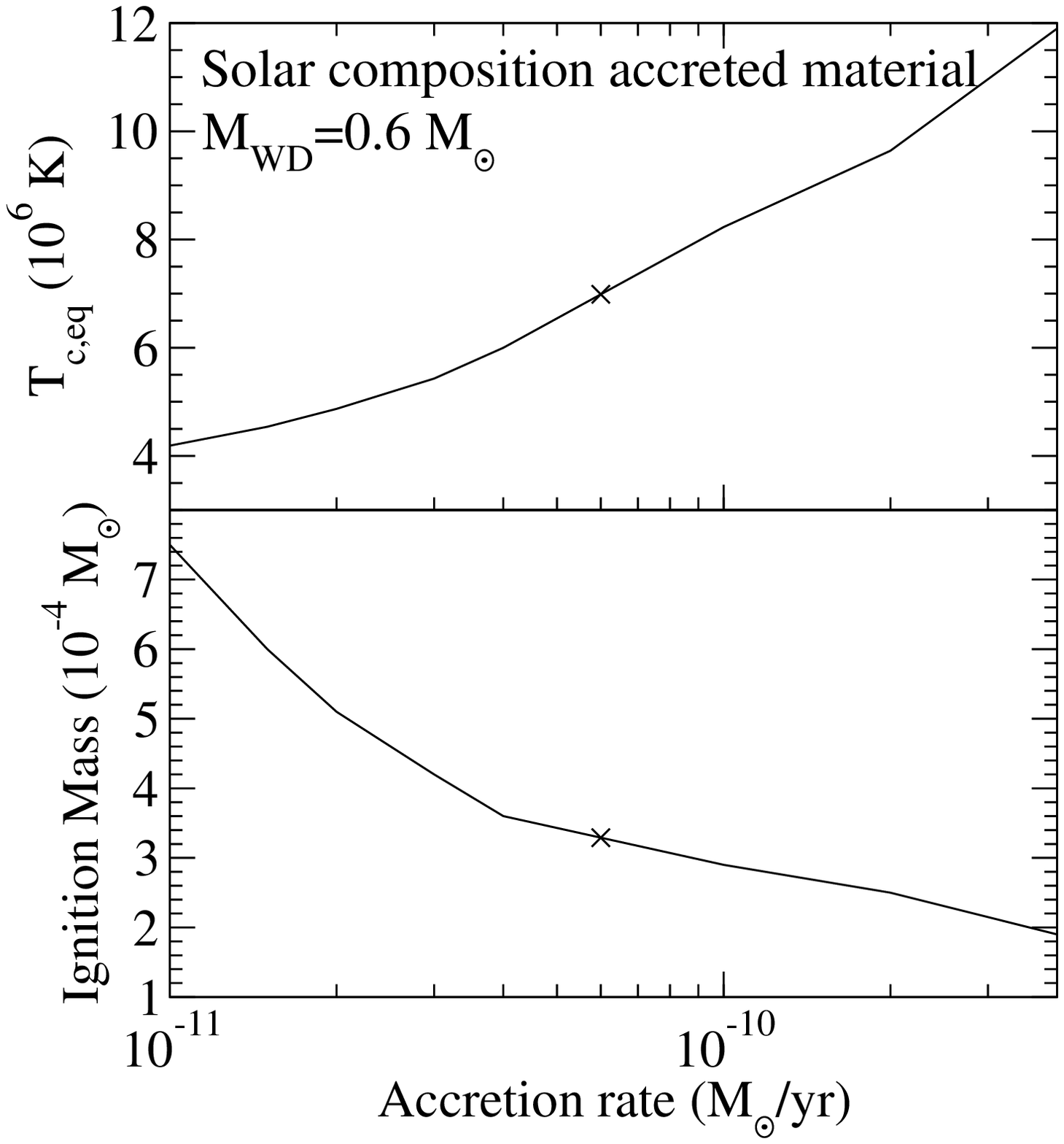}
\epsfxsize=2.32in
\epsfbox{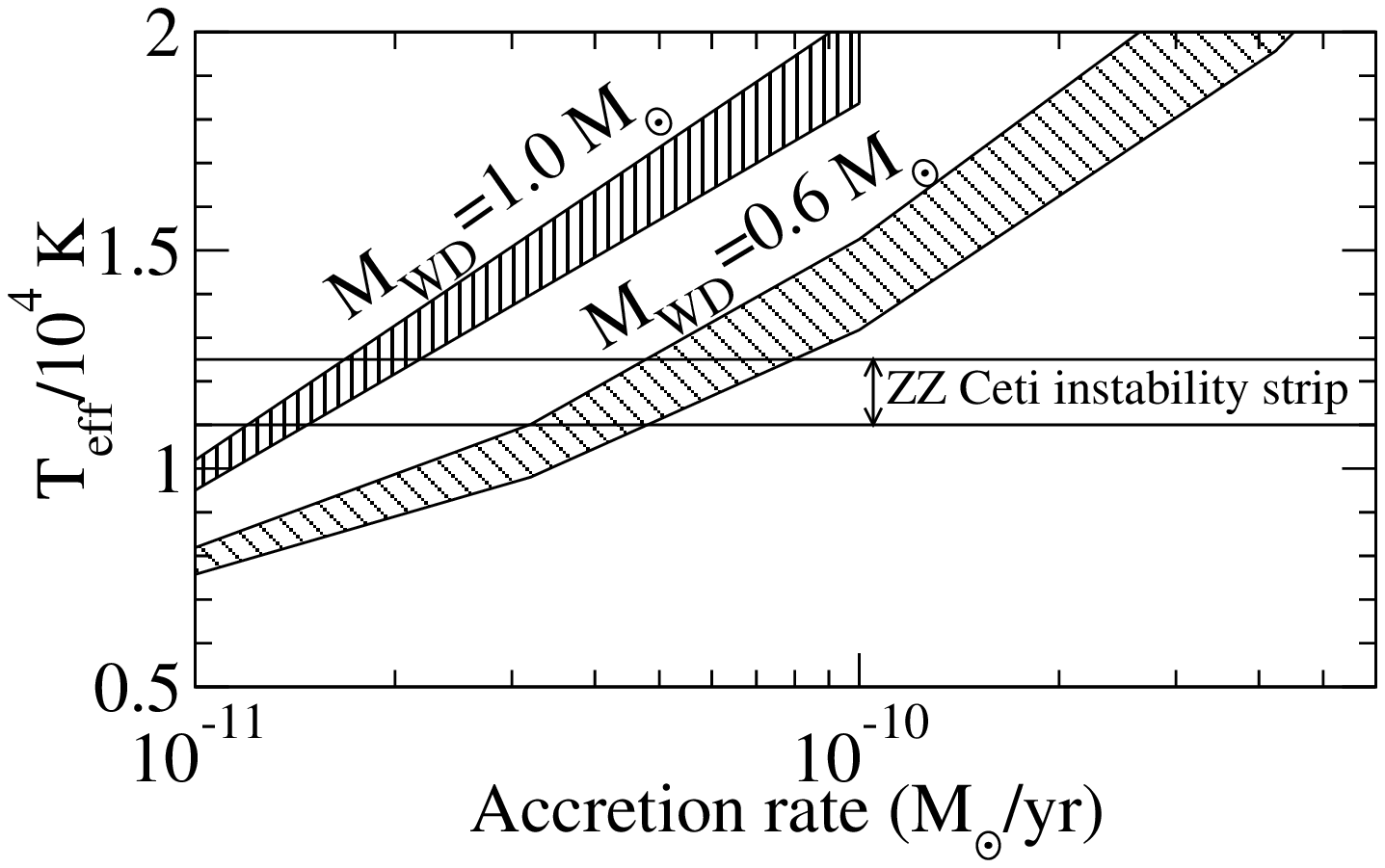}
\caption{(Left Panel) 
Properties determined in equilibrium solutions.  The $\times$
marks the transition of CN ignition from the pp cycle at low
temperatures to CNO cycle at higher temperatures. (Right Panel) 
Predicted effective temperature range over the CN cycle. }
\vspace{-.2cm}
\end{figure}

\section{Reheating the White Dwarf Core}

 Using the compressional heating estimate previously given, we now
show why we think there is enough time for a C/O WD to reach $T_{\rm
c,eq}$. The specific heat of the liquid WD core is $\approx 3k/\mu_i
m_p$, where $\mu_i\approx 14$ is the ion mean molecular weight for the
C/O core. When
the core is colder than $T_{\rm c,eq}$, it is heated by the envelope
at the rate
\begin{equation}
\label{eq:evol}
{{3k\langle \dot M \rangle}\over {2 \mu_e m_p}}(T_{\rm c,eq}-T_{\rm c})\approx
{3kM\over \mu_i m_p}{{ dT_{\rm c}\over dt}},  
\end{equation}
which we equate to the changing thermal content of a fixed mass core
(presuming all accreted matter is ejected during the CN).  This gives
the time evolution $T_c(x)=T_{\rm in}\exp(-x)+T_{\rm
c,eq}(1-\exp(-x))$, where the time coordinate is $x=(\mu_i/2\mu_e)\int
\timav dt/M\approx 11.7 \Delta M/M$ and $\Delta M$ is the total mass
transferred. The large prefactor (due to the contrast in the $\mu$'s)
allows the core temperature to match $T_{\rm c,eq}$ to within 5
percent when $\Delta M\approx 0.25 M\approx 0.15 M_\odot$, which is
easily satisfied by CV evolution. Disequilibrium can occur if the
accreting WD is mostly helium, as then $\mu_i=4$ and the specific heat
is large. 

\section{Inferring $\timav$ from $T_{\rm eff}$} 

 The right panel of Figure 1 shows our predicted $T_{\rm eff}$ range
during the $\sim 10^6 $ year CN cycle as a function of $\timav$ for WD
cores with $T_{\rm c,eq}$. Clearly our predictions are in the observed
range noted earlier. We also show on this plot the typical $T_{\rm
eff}$ range of the non-radial ZZ Ceti pulsators, making it clear that
the discovery of more such systems is expected. 
Comparing our calculations to DN systems below the period gap,
$\timav\approx 10^{-10}M_\odot\ \rm yr^{-1}$ imply that the WD masses
are in the range $0.6$-$1.0M_\odot$. This agrees with the expectation
from \cite{kolb}, who find $\timav\approx 5\times 10^{-11} M_\odot \
{\rm yr}^{-1}$ at an orbital period of 2 hours presuming angular
momentum losses from gravitational radiation. Above the period gap,
the observed $T_{\rm eff}$ is higher, and we find $\timav\approx
10^{-9}M_\odot\ \rm yr^{-1}$ \cite{towbil}. This $\timav$ contrast
above and below the period gap qualitatively confirms the ``standard''
model of CV evolution with disrupted magnetic braking (see \cite{how}
for an updated discussion).

\section{Open Issues}

 There remain many unanswered questions that we hope to address. Our
current modeling efforts are spherically symmetric and we are often
asked if disk accretion can make a difference. We think not, since the
depth of the matter when it ignites is so large that it should have
covered the whole star by that time. 
Detailed comparisons to nearby DN with accurate $T_{\rm
eff}$'s need to be undertaken by us in conjunction with the
observers. In addition, the current $T_{\rm eff}$ measurement of GW
Lib \cite{szkody} puts it outside the ZZ Ceti strip for
$M=0.6M_\odot$, reminding us that we have alot to learn about
non-radial oscillations in these very perturbed WDs.

 This work was supported by the NSF under Grants PHY99-07949,
AST01-96422, and AST02-05956. Support for this work was provided by
NASA through grant AR-09517.01-A from STScI, which is operated by
AURA, Inc., under NASA contract NAS5-26555. 
L. B. is a Cottrell Scholar of the Research Corporation and D. T. is
an NSF graduate fellow.


\begin{thebibliography}{99}

\bibitem{alt}Althaus, L. G., \& Benvenuto, O. G. 1997, ApJ, 477, 313 
\bibitem{how}Howell, S. B., Nelson, L. A., \& Rappaport, S. 2001, ApJ,
550, 897
\bibitem{kolb}Kolb, U. \& Baraffe, I. 1999, MNRAS, 309, 1034 
\bibitem{sal} Salaris, M., et al. 2000, ApJ, 544, 1036 
\bibitem{sion} Sion, E. M. 1999, PASP, 111, 532
\bibitem{szkody}Szkody, P., G\"ansicke, B. T., Howell, S. B. \& Sion,
E. M., 2002, ApJ, 575, L79 
\bibitem {towbil} Townsley, D. M., \& Bildsten, L. 2002, in 
``The Physics of Cataclysmic Variables and Related Objects'' 
ed.\ B. T. G\"ansicke et. al. (San Francisco: ASP),
p. 31. 
\bibitem {tow02} Townsley, D. M. \& Bildsten, L. 2002, ApJ, 565, L35
\bibitem{zyl} van Zyl, L. et al. 2000, Baltic Astronomy, 9, 231

\end{thebibliography}
\end{document}